\documentclass[journal=jacsat,manuscript=article]{achemso}
\usepackage{achemso}
\usepackage[version=3]{mhchem}
\usepackage[english]{babel}
\usepackage{filecontents}
\usepackage{graphicx}
\usepackage{courier}
\usepackage{amssymb,amsmath,amsthm,mathrsfs,comment,afterpage,accents}
\usepackage{mathtools}
\usepackage{physics}
\usepackage{mathrsfs}
\usepackage{courier}
\usepackage[table,xcdraw]{xcolor}
\usepackage{subdepth}
\usepackage{amsmath}
\usepackage{booktabs}
\setkeys{acs}{articletitle = true}
\makeatletter
\def\@dotsep{4.5}
\makeatother
\usepackage{dcolumn}
\usepackage{bm}
\renewcommand\vec\mathbf
\newcommand\mat\mathbf
\setkeys{acs}{maxauthors = 0}

\newcommand{\joonho}[2]{{\textcolor{red}{}}{\textcolor{black}{#2}}}
\newcommand{\bill}[1]{{\textcolor{black} {#1}}}

\newcommand{\insertnew}[1]{{\textcolor{black} {#1}}}
\newcommand{\revision}[1]{{}{\textcolor{black}{#1}}}

\title{Generalized Unitary Coupled Cluster Wavefunctions for Quantum Computation}
\author{Joonho Lee}
\altaffiliation{Authors contributed equally to this work.}
\email{linusjoonho@gmail.com}
\author{William J. Huggins}
\altaffiliation{Authors contributed equally to this work.}
\email{wjhuggins@gmail.com}
\author{Martin Head-Gordon}
\email{mhg@cchem.berkeley.edu}
\author{K. Birgitta Whaley}
\email{whaley@berkeley.edu}
\affiliation{
Department of Chemistry, University of California, Berkeley, California 94720, USA
Chemical Sciences Division, Lawrence Berkeley National Laboratory, Berkeley, California 94720, USA
}

\begin{document}
\maketitle
\newpage
\begin{abstract}
We introduce a unitary coupled-cluster (UCC) ansatz 
termed $k$-UpCCGSD that is based on a family of sparse generalized doubles operators which provides an affordable and systematically improvable unitary coupled-cluster wavefunction suitable for implementation on a near-term quantum computer. $k$-UpCCGSD employs $k$ products of the exponential of pair coupled-cluster double excitation operators (pCCD), together with generalized single excitation operators. We compare its performance in both efficiency of implementation and accuracy with that of the generalized UCC ansatz employing the full generalized single and double excitation operators (UCCGSD), as well as with the standard ansatz employing only single and double excitations (UCCCSD).
$k$-UpCCGSD is found to show the best scaling for quantum computing applications, requiring a circuit depth of $\mathcal O(kN)$, compared with $\mathcal O(N^3)$ for UCCGSD and $\mathcal O((N-\eta)^2 \eta)$ for UCCSD where $N$ is the number of spin orbitals and $\eta$ is the number of electrons. 
We analyzed the accuracy of these three ans{\"a}tze by making classical benchmark calculations on the ground state and the first excited state of H$_4$ (STO-3G, 6-31G), H$_2$O (STO-3G), and N$_2$ (STO-3G), making additional comparisons to conventional coupled cluster methods. The results for ground states show that  $k$-UpCCGSD offers a good tradeoff between accuracy and cost, achieving chemical accuracy for lower cost of implementation on quantum computers than both UCCGSD and UCCSD.
UCCGSD is also found to be more accurate than UCCSD, but at a greater cost for implementation.
Excited states are calculated {with an} orthogonally constrained variational quantum eigensolver approach. This is seen to generally yield less accurate energies than for the corresponding ground states.
We demonstrate that using a specialized multi-determinantal reference state constructed from classical linear response calculations allows these excited state energetics to be improved.
\end{abstract}
\newpage

\section{Introduction}
Quantum computing promises to provide access to a new set of computational
primitives that possess profoundly different limitations from those available
classically.
It was shown early on that quantum phase estimation (QPE) provides an
exponential speed-up over the best ``currently'' known classical algorithms for determining
the ground state of the molecular Hamiltonian.\cite{aspuru2005simulated} However, the use
of this approach is believed to require large, error-corrected, quantum
computers to surpass what is possible classically~\cite{reiher2017elucidating, babbush2018encoding}.
A more promising path to \joonho{exploring}{pursuing} such ``quantum supremacy''\cite{boixo2018characterizing,harrow2017quantum} in the context of
quantum chemistry on near-term quantum
devices is a quantum-classical hybrid algorithm that is referred to as the variational quantum
eigensolver (VQE)\cite{peruzzo2014variational}. Interested readers are referred to a more extensive review in Ref.~\citenum{McArdle2018}.

Unlike phase estimation, VQE requires only a short coherence time.
This hybrid approach uses a quantum computer to prepare and
manipulate a parameterized wavefunction, and {embeds this in} a classical optimization
algorithm to minimize the energy of the state as measured on the quantum computer, i.e., 
\begin{equation}
E = \text{min}_{\theta} \langle\psi(\theta)|\hat{H}|\psi(\theta)\rangle,
\label{eq:vqe}
\end{equation}
where $\theta$ denotes the set of parameters specifying the quantum circuit required to prepare the state $\ket{\psi}$.
From a quantum chemistry perspective, there are 
two key attractive
aspects of the VQE framework: 
\begin{enumerate}
\item The evaluation of the energy of a wide class of wavefunction ans{\"a}tze which are
  exponentially costly classically \insertnew{(with currently known algorithms)} requires only state preparation and measurement of Pauli operators, both of which can be carried out on a quantum processor in polynomial time. These
wavefunction ans{\"a}tze include unitary coupled-cluster 
(UCC) wavefunctions,~\cite{evangelista2011alternative, peruzzo2014variational} the deep multi-scale entanglement renormalization ansatz (DMERA),~\cite{kim2017robust} a Trotterized version of adiabatic state preparation (TASP),~\cite{wecker2015progress} \revision{the qubit coupled cluster approach (QCC),\cite{ryabinkin2018qubit}} and various low-depth quantum circuits inspired by the specific constraints of physical devices currently available.~\cite{kandala2017hardware}

\item On a quantum processor, efficient evaluation of the magnitude of the overlap between two
states is possible even when two states involve exponentially many determinants. Classically, this is a distinct feature only of tensor network~\cite{stoudenmire2012studying} and variational Monte  Carlo \cite{Cyrus1998} approaches.
However on a quantum computer, any states that can be efficiently prepared will also possess this advantage.
\end{enumerate}

Given the recent progress \joonho{}{and near-term prospects} in quantum computing
hardware, and the uniqueness of these capabilities, it is interesting to explore these two aspects from a quantum chemistry perspective and this constitutes the major motivation of this work.

The remainder of this paper is organized as follows. 
(1) We review existing UCC ans{\"a}tze in the context of traditional coupled cluster theory, focusing in particular on unitary extensions of the generalized
coupled-cluster ansatz of Nooijen.\cite{Nooijen2000}  We then present a new ansatz, referred to as $k$-UpCCGSD, that uses $k$ products of the exponential of distinct pair coupled-cluster double excitation operators, together with generalized single excitation operators.
We show that this ansatz is more powerful than previous unitary extensions of coupled-cluster, achieving a significant reduction in scaling of circuit depth relative to both straightforward unitary extensions of generalized UCC (UCCGSD) and conventional UCC with single and double excitations (UCCSD).
(2) We analyze options for variational optimization of excited states that are subject to orthogonalization constraints with a previously variationally optimized ground state.~\cite{higgott2018variational} 
We explore several distinct options and  make an analysis of the possible errors encountered when using such a variational approach.  We show that these excited state energies can be significantly improved
by using a different reference state for the excited state variational calculation, specifically, by using single excitation reference states. 
(3) We undertake a systematic analysis of the resource requirements for realization of these UCC ans{\"a}tze on a quantum computer, relevant to preparation of initial states of molecules for both QPE and VQE computations. Our resource analysis focuses on the scaling of gate count, circuit depth, and spatial resources with size of the quantum chemistry calculation.   We find that the $k$-UpCCGSD ansatz exhibits a linear dependence of circuit depth (a measure of the computational time that we define explicitly below) on the number of spin-orbitals $N$, with higher order polynomial dependence obtained for both UCCGSD and UCCSD.
(4) To assess the accuracy of the new ansatz, we undertake benchmarking calculations on a classical computer for ground and first excited states of three small molecular systems, namely \ce{H4} (STO-3G, 6-31G), \ce{H2O} (STO-3G), and \ce{N2} (STO-3G), making additional comparisons to conventional coupled cluster methods as relevant. Detailed analysis of potential energy curves for ground and excited states of all three species shows that $k$-UpCCGSD ansatz offers the best trade-off between low cost and accuracy.
(5) We conclude with a summary and outlook for further development of unitary coupled cluster ans{\"a}tze for efficient implementation of molecular electronic states in quantum computations.

\section{Theory}
We shall use $i,j,k,l,\cdot\cdot\cdot$ to index occupied orbitals, $a,b,c,d,\cdot\cdot\cdot$ to index unoccupied (or virtual) orbitals, and $p,q,r,s,\cdot\cdot\cdot$ to index either of these two types of orbitals. The indices will denote spin-orbitals unless mentioned otherwise. We use $N$ to denote the number of spin-orbitals and $\eta$ to denote the number of electrons.

\subsection{Coupled-Cluster Theory}
In this section, we first briefly review traditional coupled cluster (CC) theory 
and unitary CC (UCC). We shall then draw connections between an existing body of work on variants of coupled cluster theory and a recently described wavefunction ansatz {for VQE,}\cite{wecker2015progress} before proposing a novel ansatz also motivated by previous work in quantum chemistry. 
\revision{We note that in the quantum information literature it is customary to use UCC to denote the unitary version of restricted CC, in contrast to the quantum chemistry literature where UCC generally refers to unrestricted CC. We follow the quantum information convention in this paper.}

\subsubsection{Traditional Coupled Cluster}
Traditional CC
is a successful wave function method used for
treating correlated systems in quantum chemistry.\cite{Cizek1966,Cizek1980,Bartlett2007} Coupled-{c}luster with singles and doubles (CCSD), i.e., where the excitations in the cluster operator \(\hat{T}\) are restricted
to singles and doubles, is suitable for treating most ``weakly-correlated'' chemical systems. 

The CCSD wave function is usually written with an exponential generator acting on a reference state,
\begin{align}
|\psi\rangle &= e^{\hat{T}}|\phi_0\rangle,
\end{align}
where for CCSD we have a cluster operator
\begin{equation}
\label{eq:T}
  \hat{T} = \hat{T}_1 + \hat{T}_2,
\end {equation}
with
\begin{align}
  \label{eq:T_singles}
  \hat{T}_1 &= \sum_{ia}t_i^a \hat{a}^\dagger_a \hat{a}_i \\
  \label{eq:T_doubles}
  \hat{T}_2 &= \frac14\sum_{ijab} t_{ij}^{ab} \hat{a}^\dagger_a \hat{a}^\dagger_b \hat{a}_j \hat{a}_i.
\end{align}
In traditional CCSD, we evaluate the energy \joonho{by making use of the following projection equation,}
{by projection of the Schr{\"o}dinger equation, $\hat{H}|\psi\rangle = E |\psi\rangle$ first with $\langle\phi_0|$:}
\begin{equation}
E \equiv \langle \phi_0|\hat{H} |\psi\rangle.
\label{eq:Eccsd}
\end{equation}
\joonho{}{We then project with $\langle\phi_\mu|$ where $\mu$ is any single ($\langle\phi_i^a|$) or double ($\langle\phi_{ij}^{ab}|$) substitution.}
The $t$-amplitudes are then obtained by solving a set of non-linear equations:
\begin{equation}
0 = \langle \phi_\mu|\hat{H} |\psi\rangle-Et_\mu,
\label{eq:tccsd}
\end{equation}
with $|\phi_\mu\rangle = \hat{t}_\mu |\phi_0\rangle$.
The cost of solving Eq. \eqref{eq:tccsd} scales as $\mathcal{O}(\eta^2(N-\eta)^4)$, where $\eta$ is the number
of electrons and $N$ is the total number of spin-orbitals possessed by the system.

It is evident from Eq. \eqref{eq:Eccsd} that the projective way of evaluating energy
is not in general variational, except in some obvious limits where CCSD is exact (e.g., for
non-interacting two-electron systems\cite{Cizek1966,Cizek1980,Bartlett2007}).
With spin-restricted orbitals, it is quite common to observe catastrophic
non-variational failure of CCSD when breaking bonds or, more broadly, in the
presence of strong correlation.
This non-variational catastrophe is often attributed to the way in which traditional CCSD
parametrizes quadruples (i.e., $\hat{T}_2^2/2!$)\cite{Paldus1984,Piecuch1995,Small2012,Small2014,Lee2017,Lee2018} and searching for solutions to this problem without
increasing the computational cost is an active area of research.\cite{Small2012,Small2014,Lee2017,Lee2018} Unfortunately, attempting to avoid this
breakdown by variationally evaluating the energy of a CC wave function
leads to a cost that scales exponentially with system size.
\subsubsection{Unitary CC}
A simple approach to avoid the non-variational catastrophe on a quantum computer is to employ a unitary CC (UCC) wavefunction,\cite{Kutzelnigg1982,Kutzelnigg1983,Kutzelnigg1984,Kutzelnigg1985,Taube2006}
\begin{align} \label{eq:UCC}
    |\psi\rangle = e^{\hat{T} - \hat{T}^\dag}|\phi_0\rangle,
\end{align}
where for the case of UCCSD, $\hat{T}$ is defined as in Eqs. \eqref{eq:T} - \eqref{eq:T_doubles}.
We can then evaluate the energy in a variational manner,
\begin{equation}
E(\{t_i^a\},\{t_{ij}^{ab}\}) \equiv \frac{\langle\psi|\hat{H}|\psi\rangle}{\langle\psi|\psi\rangle},
\label{eq:Euccsd}
\end{equation}
using the standard VQE approach\cite{peruzzo2014variational,mcclean2016theory,McArdle2018} that is summarized later in this work. UCC has a long
history in electronic structure for quantum chemistry, with a number of theoretical works dedicated to the
approximate evaluation of Eq. \eqref{eq:Euccsd} within a polynomial amount of
time,\cite{Kutzelnigg1982,Kutzelnigg1983,Kutzelnigg1984,Kutzelnigg1985,Taube2006}
since the approach appears to scale exponentially if implemented exactly using a classical computer. UCC is more \joonho{flexible}{robust} than
traditional CC, due to the fact that the unitary cluster operator involves not only excitation
operators ($\hat{T}$) but also de-excitation operators ($\hat{T}^\dagger$). Nevertheless,
the single reference nature of Eq. \eqref{eq:UCC} can still lead to difficulties when treating
strongly correlated systems on classical computers. \revision{This was investigated in Ref. \citenum{harsha2018difference} for the Lipkin Hamiltonian.}

Unlike a classical computer, a quantum computer can efficiently employ a UCC
wavefunction\insertnew{, even with a complicated multi-determinantal reference state,} since both preparation of the state and evaluation of its expectation values can be carried out using
resources that scale polynomially with system size and number of electrons.\cite{peruzzo2014variational,mcclean2016theory} For UCC with singles and doubles (UCCSD), one
must implement a Trotterized version of the exponentiated cluster operator, with \(\mathcal{O}((N-\eta)^2\eta^2)\) terms, where each term
acts on a constant number of spin-orbitals.

\subsubsection{Generalized CC}
In the early 2000's, there was an active debate on the question of whether the exact ground state wavefunction of an electronic Hamiltonian can always be represented by a general two-body cluster expansion. Motivated by earlier work of Nakatsuji,\cite{Nakatsuji1976} Nooijen conjectured \cite{Nooijen2000} that it is possible to express an exact ground state of a two-body Hamiltonian as
\begin{equation}
|\psi\rangle = e^{\hat{T}}|\phi_0\rangle,
\end{equation}
where
\begin{align}
\label{eq:generalized_T}
\hat{T} &= \hat{T}_1 + \hat{T}_2\\
&=
\frac12\sum_{pq} t_{p}^q \hat{a}_{q}^\dagger\hat{a}_p
+\frac14\sum_{pqrs} t_{pq}^{rs} \hat{a}_{r}^\dagger\hat{a}_{s}^\dagger\hat{a}_q\hat{a}_p.
\end{align}
This yields an exponential ansatz with a number of free parameters, the \(t_{p}^q\) and \(t_{pq}^{rs}\) values, that is equal to the
number of parameters in the Hamiltonian.
Here the single and double ``excitation" terms do not distinguish between
occupied and unoccupied orbitals and they are therefore called ``generalized''
singles and doubles (GSD). Although early work showed that the numerical
performance of the resulting wavefunction was promising, {the conjecture of Ref.\citenum{Nooijen2000}} has
been the subject of an active debate and was later
disproved.\cite{Nakatsuji2000,VanVoorhis2001,Piecuch2003,Davidson2003,Ronen2003,Mazziotti2004,Nakatsuji2004,Mukherjee2004}

\subsection{Generalized Unitary CC}
We explore here a generalized form of the UCC wavefunction introduced in the VQE literature.\cite{peruzzo2014variational} Our approach uses the generalized excitations of of Nakatsuji and Nooijen described above in the ansatz
\begin{equation}
|\psi\rangle = e^{\hat{T} - \hat{T}^\dagger}|\phi_0\rangle,
\end{equation}
{with \(\hat{T}\)} the cluster operator from Eq. \eqref{eq:generalized_T}. We shall term this ansatz UCCGSD.  A unitary version of coupled cluster with generalized singles and doubles was first mentioned in Nooijen's paper,\cite{Nooijen2000} but has never been thoroughly studied classically without
making an approximation to the energy evaluation.

We note that a similar approach to defining a UCC ansatz by relating the terms in the
Hamiltonian to generalized singles and doubles operators has appeared recently in the quantum computing
literature,~\cite{wecker2015progress} where the performance of a Trotterized version of such a UCCGSD on small hydrogen chains and equilibrium geometry molecular systems has been characterized.
As we shall show explicitly later in this
work, the UCCGSD wavefunction is far more robust and accurate than the simpler UCCSD wavefunctions for the chemical applications considered here.

\subsubsection{Unitary Pair CC with Generalized Singles and Doubles Product Wavefunctions}
The method of pair coupled-cluster double excitations (pCCD),\cite{Stein2014} also known as AP1roG,\cite{Limacher2013} extends
a widely used quantum chemistry method known as generalized valence-bond perfect-pairing (GVB-PP) \cite{Goddard1973}.
pCCD is less prone than spin-restricted CCSD (RCCSD) to a non-variational failure when breaking bonds, 
despite the fact that it is computationally much simpler than RCCSD.
pCCD is a coupled cluster wavefunction with a very limited number of doubles amplitudes (containing only the two body excitations that move a pair of electrons from one spatial orbital to another),
\begin{equation}
\hat{T}_2 = \sum_{ia} t_{i_\alpha i_\beta}^{a_\alpha a_\beta}
\hat{a}^\dagger_{a_\alpha} \hat{a}^\dagger_{a_\beta}
\hat{a}_{i_\beta} \hat{a}_{i_\alpha},
\label{eq:Tpccd}
\end{equation}
where the summation runs over occupied and unoccupied spatial orbitals. pCCD is capable of breaking a single-bond qualitatively correctly, but fails to break multiple bonds. 
Orbital optimization of pCCD wavefunctions includes the important effects of the single excitations in a UCC wavefunction.
\joonho{}{In exchange for its high computational efficiency and reduced incidence of non-variationality, pCCD has other disadvantages: it loses invariance to unitary transformation within the occupied-occupied and virtual-virtual subspaces present in CCD, and it does not recover the dynamic correlation that CCD has.}

We define the unitary pCCSD (UpCCSD) wavefunction to have the full singles operator as in Eq. \eqref{eq:T_singles} together with the unitary doubles operator of Eq. \eqref{eq:Tpccd}. We show below in the analysis of the quantum resource requirements that the circuit depth (time complexity) of preparing a UpCCSD state on a quantum computer scales linearly with the system size as quantified by the number of spin-orbitals. However, our initial exploration of UpCCSD yielded errors in the absolute energies that were generally larger than the threshold for chemical accuracy. We therefore improve this wavefunction by the following two modifications: (i) we use the generalized singles and doubles operators employed in
Refs. \citenum{Nakatsuji1976}, \citenum{Nooijen2000}, and (ii) we take a product of a total of $k$ unitary operators to increase the flexibility of the wavefunction. We shall refer to this model as $k$-UpCCGSD.

Formally, $k$-UpCCGSD is defined in the following manner. For a \joonho{given}{chosen} integer $k$,
\begin{equation}
|\psi\rangle
=
\Pi_{\alpha = 1}^{k}\left(e^{\hat{T}^{(\alpha)}- {\hat{T}^{(\alpha)}}^\dagger}\right)
|\phi_0\rangle,
\end{equation}
where each $\hat{T}^{(k)}$ contains an independent set of variational parameters
(i.e., {the} singles and paired doubles amplitudes, the \(t_{p}^q\)'s and the \(t_{p_\alpha p_\beta}^{q_\alpha
  q_\beta}\)'s respectively). Since the doubles operator in UpCCGSD
is very sparse, the circuit depth required to prepare a $k$-UpCCGSD state still scales
linearly with the system size, with a prefactor that is increased by a factor of $k$. This is
similar in spirit to other recently proposed low depth
ans{\"a}tze~\cite{dallaire2018low} and also to the repeated independent variational steps of the Trotterized adiabatic state preparation approach\cite{wecker2015progress} but, to our knowledge, this form of wavefunction has never
been explored in either classical or quantum computational electronic structure calculations for quantum chemistry.


\subsection{Excited State Algorithms}
\subsubsection{Previous Work}
Obtaining excited states under the variational quantum eigensolver (VQE)
framework has attracted considerable interest recently due to the substantial progress made in {experimental} realization of ground state VQE simulations\cite{peruzzo2014variational,santagati2018witnessing,omalley2016scalable,shen2017quantum,colless2017robust,kandala2017hardware} Algorithms proposed to extend this hybrid approach to excited states include the quantum
subspace expansion (QSE) algorithm~\cite{mcclean2017hybrid},  the folded spectrum (FS)
method~\cite{peruzzo2014variational}, the witnessing eigenstates (WAVES) strategy~\cite{santagati2018witnessing}, and a method based on penalizing
overlap with an approximate ground state~\cite{higgott2018variational,endo2018discovering}. We shall
refer to the last of these as orthogonally constrained VQE (OC-VQE). 

The QSE method is motivated by a linear-response approach: it samples the
Hamiltonian matrix elements in the linear response space of a ground state wave
function and diagonalizes it to obtain an excitation spectrum. A major drawback
of this method is an obvious steep increase in the number of measurements after
the ground state VQE calculation, since every matrix element needs to be sampled.
Furthermore, QSE suffers from the well-known problem of linear-response methods,
that is, it can only describe excited states that are within a small
perturbation of a given ground state. However, the proper description of chemically relevant excited states \joonho{might require inclusion of a much}{sometimes requires inclusion of a} higher order of excitations. A classic example of this is the dark low-lying
excited state of butadiene, which requires that the linear response space include quadruple excitations in order to obtain a converged result.\cite{Watson2012}

The FS method is closely related to the variance minimization algorithm widely
used in the quantum Monte Carlo community:\cite{Choi1970}
\begin{equation}
E(\omega) = \text{min}_{\theta} \langle\psi(\theta)|(\hat{H}-\omega)^2|\psi(\theta)\rangle.
\label{eq:var}
\end{equation}
One advantage of this algorithm over the WAVES and OC-VQE algorithms is its ability to
target a state whose energy is the closest to a preset $\omega$, as in Eq.
\eqref{eq:var}. Although this ability to variationally target specific excited states is very desirable, the algorithm inherently involves the evaluation of a quadratic term
in $\hat{H}$, which greatly increases the number of Hamiltonian terms. Due to its steep
scaling, \(\mathcal{O}(N^8)\) in a standard gaussian basis set, application of the FS method (if possible) is likely to be limited to very small systems.

The WAVES algorithm relies on the ability of a quantum computer to efficiently
perform time evolution conditioned on the state of a control qubit.\cite{santagati2018witnessing}  The protocol
applies single qubit tomography to the first qubit of the state
\(\frac{1}{\sqrt{2}}\ket{0}\otimes \ket{\psi} + \frac{1}{\sqrt{2}}\ket{1}
\otimes e^{-i\hat{H}t}\ket{\psi}\), for a given input state \(\ket{\psi}\) and
time \(t\). The reduced density matrix of the control qubit describes a pure
state if and only if \(\ket{\psi}\) is an eigenstate of the Hamiltonian, or a superposition of degenerate eigenstates. Using this idea, it is possible to
variationally target excited states (although not specific energies as is possible with the FS method), by varying the parameters of the trial state to maximize the purity of the measured single qubit state. This advantage is offset by the requirement
that the quantum computer must implement a controlled version of the time
evolution operator, which imposes steep demands on the relatively noisy quantum computing devices currently available.

\subsubsection{Orthogonally Constrained VQE}
In this work we explore an alternative to the aforementioned three methods
which has the advantage that it requires roughly the same number of measurements
as the ground state VQE calculation and only a doubling of the necessary circuit
depth. \cite{higgott2018variational}
This algorithm can be naturally used with the two generalized coupled cluster wavefunction ans{\"a}tze
described above, or with any other circuit suitable for ground state VQE.
Furthermore, OC-VQE can describe excited states that lie beyond the
linear-response regime of the ground state.
The approach assumes that a circuit for the ground state wavefunction is already
available from a standard VQE calculation. One then defines an effective
Hamiltonian whose lowest eigenstate is the first excited state and whose lowest
eigenvalue is the energy of said state. 

\revision{One such choice is given by 
\begin{equation}
\hat{H}_\text{OC-VQE} = \hat{H} + \mu\ket{\psi_{0}}\bra{\psi_0},
\label{eq:ocvqe}
\end{equation}
where \(|\psi_0\rangle\) is the ground state wavefunction and the second term constitutes a level shift operator.
For the molecular systems studied here, both the ground and first excited states are bound states (i.e., the electronic energies of these states are negative).
Under these assumptions, we can choose
\(\mu = -E_0 =-\langle\psi_0|\hat{H}|\psi_0\rangle \). \cite{higgott2018variational}
This level shift imposes an energy penalty of $\mu|\langle\psi_0|\psi_1\rangle|^2$ on any trial state $|\psi_1\rangle$ that overlaps with $|\psi_0\rangle$.
Such an energy level shift technique is commonly used in quantum chemistry to enforce constraints within a variational
framework~\cite{Andrews1991,Evangelista2013,Glushkov2016,Manby2012}. 
Similar techniques have also been used in density matrix renormalization group calculations.~\cite{stoudenmire2012studying}
Minimizing the expectation value of $\hat{H}_\text{OC-VQE}$ with respect to the parameters in $\ket{\psi_1}$ defines this first OC-VQE procedure. 
}

The choice of effective Hamiltonian in Eq. \eqref{eq:ocvqe} is not unique. We have also explored the form
\begin{equation}
 \hat{H}_\text{OC-VQE}' = \left(1 -\ket{\psi_{0}}\bra{\psi_0}\right)\hat{H} \left(1 - \ket{\psi_{0}}\bra{\psi_0}\right).
 \label{eq:ocvqe2}
 \end{equation}
 Eq. \eqref{eq:ocvqe} and Eq. \eqref{eq:ocvqe2} are identical if and only if $\ket{\psi_0}$ is an eigenstate of $\hat{H}$ with an eigenvalue $E_0$. If we choose $\mu = \infty$, the two \revision{approaches} yield the same first excited state for a given approximate ground state \(\ket{\psi_0}\). 
Both Eqs.~\eqref{eq:ocvqe} and \eqref{eq:ocvqe2} minimize the trial energy in the orthogonal complement space of $|\psi_0\rangle$, and these two different effective Hamiltonians have been interchangeably utilized in various contexts in quantum chemistry. \cite{Manby2012, Evangelista2013}
We choose to work with Eq. \eqref{eq:ocvqe} {here, since} it has a clear
implementation suitable for a near term quantum device without requiring costly
controlled unitary implementations of the state preparation circuits. 

Specifically, it is clear that OC-VQE can be effectively implemented using the
Hamiltonian of Eq.~\eqref{eq:ocvqe} so long as an efficient algorithm for measuring the
magnitude of the overlap between the ground state and a trial excited state is available. On
a classical computer, measuring the overlap between, for instance, two UCC states scales exponentially while on a quantum device this task is only polynomial scaling.\cite{higgott2018variational} We describe one implementation of the necessary
overlap calculation between two parameterized quantum states in the Quantum Resource Requirements section below,
and refer the reader to recent work by Higgott et
al.~\cite{higgott2018variational} for additional discussion on minimizing the
effect of errors on this measurement.

\subsubsection{Energy Error Analysis of OC-VQE}
When an exact ground state $|\psi_0\rangle$ of \(\hat{H}\) is used to construct the effective Hamiltonian $\hat{H}_\text{OC-VQE}$ in Eq. \eqref{eq:ocvqe}, the exact ground state of $\hat{H}_\text{OC-VQE}$ yields the exact excited state of the original Hamiltonian \(\hat{H}\).
We now show that use of an approximate ground state, \(\tilde{\ket{\psi_0}}\), in the construction of $\hat{H}_\text{OC-VQE}$ will cause the excited state energy to incur an error that is similar in size to the error in the ground state energy, i.e. \(E_0 - \tilde{\bra{\psi_0}} \hat{H} \tilde{\ket{\psi_0}}\). We define the relevant excited state Hamiltonian,
\begin{equation}
\hat{\tilde{H}}_\text{exc} = \hat{H} - \tilde{E_0}\tilde{\ket{\psi_0}}\tilde{\bra{\psi_0}},
\label{eq:Hexc_approx}
\end{equation}
and consider the difference in energy between the ground states of $\hat{\tilde{H}}_\text{exc}$ and of \(\hat{H}_{exc}\) in Eq.~\eqref{eq:ocvqe}.

Writing the approximate ground state as \(\tilde{\ket{\psi_0}} =
\sqrt{1-\epsilon^2}\ket{\psi_0} + \epsilon \ket{\psi_\perp}\), where
\(\braket{\psi_0}{\psi_\perp} = 0\), we can rewrite Eq.~\ref{eq:Hexc_approx} as
\begin{equation}
  \hat{\tilde{H}}_\text{exc} = \hat{H}_\text{exc} + \hat{V}, \;\;\;\;\;\hat{V} = -\epsilon E_0 \ket{\psi_\perp}\bra{\psi_0} - \epsilon E_0 \ket{\psi_0}\bra{\psi_\perp} + \mathcal{O}(\epsilon^2).
\label{eq:Hexc_expansion}
\end{equation}
The first excited state of \(\hat{H}\), which we denote \(\ket{\psi_1}\), is by definition an approximation to the ground state of \(\hat{\tilde{H}}_\text{exc}\). Assuming that
\(\epsilon\) is small, we compute the first order correction to the energy using Eq.~\eqref{eq:Hexc_expansion}. Because \(\ket{\psi_0}\) and \(\ket{\psi_1}\) are
orthogonal, it is immediately clear that \(\ev{V}{\psi_1}\) is zero to
first order in \(\epsilon\). Therefore, the difference between the true excited
state energy, \(E_1\), and the energy given by finding the ground
state of the approximate excited state Hamiltonian, \(\hat{\tilde{H}}_\text{exc}\), is \(\mathcal{O}(\epsilon^2)\), which is on the same scale as the error in the ground state energy, \(\epsilon^2(\ev{\hat{H}}{\psi_\perp} - E_0)\).

Of course, in practice, we also do not find the exact ground state energy of
\(\hat{\tilde{H}}_\text{exc}\), instead incurring an additional error in
our determination of the excited state energy from the second round of
approximate minimization. However, if we make the assumption that the VQE
procedure on \(\hat{\tilde{H}}_\text{exc}\) is carried out well enough (and the ansatz is flexible enough) to yield an approximate ground state which is \(\epsilon_1\) away from the true ground state of \(\hat{\tilde{H}}_\text{exc}\), then our overall error in the energy will be \(\mathcal{O}(\epsilon^2 + \epsilon_1^2)\).

\section{Quantum Resource Requirements}

To assess the benefits of unitary coupled cluster theory for quantum computation it is important to quantify the cost of both state preparation and measurement needed to use these states on quantum processors.
Our presentation here addresses the resources required for state preparation for a general quantum computation - we refer the reader to prior work for additional details {specific to} measurement in the VQE hybrid implementation~\cite{mcclean2016theory}.
This resource analysis requires an accounting of the number of quantum gates (``gate count'' or ``circuit size''), the time required to implement them, and the number of qubits on which they act. 
{We shall take the total gate count to be determined by the number of two-qubit gates.}
In general, the relationship between the gate count and the number of sequential time steps required to implement them when parallelization is taken into account, the ``circuit depth,'' will depend on the architectural details of the quantum processor. For many applications in quantum chemistry optimal results can nevertheless be obtained with minimal assumptions.\cite{kivlichan2018quantum,motta2018low}

We now present the implementation details necessary for evaluating the scaling of our proposed ans{\"a}tze with respect to the numbers of spin-orbitals and electrons represented by the state.
Our presentation {here} addresses the resources required for a general quantum computation - we refer the reader to prior work for additional details {specific to} the VQE hybrid implementation.\cite{mcclean2016theory}

In order to treat the UCC ansatz on a quantum computer,
it is necessary to map~\cite{ortiz2001quantum, seeley2012bravyi,bravyi2017tapering}
the reference state and the exponentiated cluster operator from a
Hilbert space of \(N\) fermionic spin-orbitals to a collection of quantum
gates acting on \(N\) qubits. Therefore, the \joonho{spatial}{qubit} resource requirement is linear in the number of spin-orbitals.  
For a UCC ansatz, the total gate count would be na{\"i}vely expected to be {lower bounded by the number of cluster amplitudes $t^q_p$ and $t^{rs}_{ps}$, possibly with additional overhead deriving from the mapping to fermionic modes and the limited connectivity of a real device}. Regarding the former, while
{the Jordan-Wigner transformation} allows the representation of fermionic
creation and annihilation operators in terms of products of single qubit Pauli
operators in a way that properly encodes the canonical commutation relations,\cite{ortiz2001quantum}  
 {direct} application of this transformation maps the fermionic operators
{acting} on individual spin-orbitals to qubit operators that act non-locally on
\(\mathcal{O}(N)\) qubits, leading to a corresponding overhead for the circuit
depth. 
However, recent work in Refs.~\citenum{motta2018low} and \citenum{OGorman2018}
describes procedures for implementing a Trotter step of unitary coupled cluster in a manner
that not only entirely eliminates this Jordan-Wigner overhead, but also allows
for the parallel implementation of individual exponentiated terms from the cluster operator {on a linearly connected array of qubits}.  
We note that a practical implementation of UCC relies on approximating
\(e^{\hat{T} - \hat{T}^\dagger}\) by a small number of 
Trotter steps, which
leads to ans{\"a}tze that are not exactly equivalent to the ones considered in our numerical
calculations. Nevertheless, it has been demonstrated that \bill{the variational optimization of} as few as one Trotter
steps \bill{of UCC} can yield highly accurate quantum chemical calculations.\cite{barkoutsos2018quantum}

Energy
measurement and wavefunction optimization in the VQE framework both require repeated state
preparation to overcome the statistical nature of the measurement process.\cite{peruzzo2014variational,mcclean2016theory,mcclean2016theory}
Therefore, in analyzing the asymptotic time complexity for quantum computation of the approaches considered here,
we focus on the cost of state preparation
as quantified by the gate count and the circuit depth required for a fixed number of Trotter steps.  
Generally, we expect a practical benefit from minimizing both the number of free parameters that must be optimized (i.e., the cluster amplitudes) and the circuit depth.

The scaling of the circuit depth was derived here by assuming the maximum possible parallelization of terms in the cluster operator that act on distinct spin-orbitals and neglecting the Jordan-Wigner overhead.\cite{OGorman2018}
Within this approach it is then clear that the $k$-UpCCGSD ansatz allows reduction of the circuit depth from the gate count by a factor of $N$, since the doubles pairs {may be grouped into \(\mathcal{O}(N)\) sets of \(\mathcal{O}(N)\) terms, each of which acts on distinct spin-orbitals and} can the \(\mathcal{O}(N)\) sets can therefore be executed in parallel.
We note that the results can also be obtained by using the procedure in Ref.~\citenum{motta2018low} without additional numerical truncation. The resulting asymptotic scaling of gate count and circuit depth with respect to both the number of spin-orbitals $N$ and electrons $\eta$   is shown in Table 1 for all three unitary ans{\"a}tze. Specific values for the numbers of cluster amplitudes used for the individual molecules for which benchmarking studies are performed will be shown in Table \ref{tab:final} in the results section.

\begin{table}[h]
\resizebox{.5\textwidth}{!}{%
\begin{tabular}{@{}lll@{}}
\toprule
{\color[HTML]{000000} Method} & {\color[HTML]{000000} Gate Count}             & {\color[HTML]{000000} Circuit Depth}                       \\ \midrule
{\color[HTML]{000000} UCCSD}  & {\color[HTML]{000000} \(\mathcal{O}((N-\eta)^2 \eta^2)\)} & {\color[HTML]{000000} \(\mathcal{O}((N-\eta)^2\eta)\)} \\
{\color[HTML]{000000} UCCGSD} & {\color[HTML]{000000} \(\mathcal{O}(N^4)\)}               & {\color[HTML]{000000} \(\mathcal{O}(N^3)\)}             \\
$k$-UpCCGSD                     & \(\mathcal{O}(k N^2)\)                                    & \(\mathcal{O}(k N)\)                                    \\ \bottomrule
\end{tabular}}
\caption{Resources required for preparing the three classes of UCC wavefunctions UCCSD, UCCGSD, and $k$-UpCCGSD, on a quantum device using a fixed number of Trotter steps. The gate count refers to the total number of quantum gates. 
The circuit depth is the number of sequential steps allowing for quantum gates acting on neighboring qubits to be executed in parallel (see text for details). $\eta$ denotes the number of electrons and \(N\) the number of spin-orbitals in the active space for a given molecule.  \(k\) denotes the number of products in the $k$-UpCCGSD wavefunction.
}
\label{tab:resource_count}
\end{table}

\subsection{Quantum implementation of Overlap Measurements}

In order to implement the excited state algorithm used
this work, Eq. \eqref{eq:ocvqe}, it is necessary to estimate
not only the expectation value of the energy, but also \(|\langle \psi_0 |
\psi_1(\theta)  \rangle|^2\), where \(| \psi_0 \rangle\) is a parameterized guess for
the ground state wavefunction and \(|\psi_1 (\theta)\rangle\) is the excited
state ansatz. Allow \(\hat{U}_1\) to be the quantum circuit that generates
\(|\psi_1 (\theta)\rangle\) from the \(|0\rangle\) state of the qubit register, i.e.,
\(|\psi_1 (\theta)\rangle = \hat{U}_1 | 0 \rangle\). Let \(\hat{U}_0\) be the unitary
which prepares \(|\psi_0\rangle\). The circuit that applies \(\hat{U}_0^\dag\) can
be constructed simply by inverting each of the gates that compose \(\hat{U}_0\).
The quantity \(|\langle \psi_0 | \psi_1 (\theta)  \rangle|^2\) can
therefore be rewritten as \(|\langle
0 | \hat{U}_0^\dag \hat{U}_1 | 0  \rangle|^2\). This is exactly equal to the
probability that the zero state will be observed when the state \(\hat{U}_0^\dag
\hat{U}_1 | 0  \rangle\) is measured in the computational basis. {Consequently,} the
magnitude of the overlap may estimated by repeated state preparation and measurement. Because of the necessity to apply both \(\hat{U}_1\) and \(\hat{U}_0^\dagger\), these measurements require a doubling of the circuit depth compared to the other observables. However, the overall cost of the measurements required for the OC-VQE approach for quantum chemistry in a molecular orbital basis will {still} be dominated by the measurement of the \(\mathcal{O}(N^4)\) terms in the original Hamiltonian.

\section{Benchmark implementations on a Classical Computer}
\subsection{Computational Details}
All the full configuration interaction (FCI) calculations needed to benchmark the demonstration examples in this work are performed through \texttt{Psi4}\cite{Parrish2017} along with its \texttt{OpenFermion}\cite{McClean2017}
interface. All UCCSD calculations are performed with an in-house code that
uses \texttt{OpenFermion}\cite{McClean2017} together with \texttt{TensorFlow}\cite{Abadi2016} for efficient gradient evaluations.
The energy as a function of the cluster amplitudes is computed variationally as
in Eq.~\ref{eq:Euccsd} and the gradient of this function is used in
conjunction with \texttt{SciPy}'s implementation of the BFGS algorithm,\cite{SciPy} a quasi-Newton
method for optimization which does not require explicit calculation of the
Hessian. 
\insertnew{The limit of our code is about 16 spin-orbitals, which allowed us to examine various model systems presented below.
A production level code may follow the implementation of Evangelista
\cite{evangelista2011alternative}, which may facilitate 
prototyping VQE ans{\"a}tze.
}
All other calculations required for the demonstrations presented in this work
are done with the development version of \texttt{Q-Chem}.\cite{Shao2015}
\revision{All calculations were performed with the frozen core approximation applied to oxygen and nitrogen.}

\revision{
There are several possible strategies for optimizing the amplitudes of a $k$-UpCCGSD wavefunction. One attractive approach is to
optimize only one set of amplitudes in $\hat{T}^{(k)}$, while fixing all the amplitudes associated with a $(k-1)$-UpCCGSD wavefunction. 
This has the potential benefit of reducing the extra computational cost for optimization of more amplitudes as the index $k$ is increased. 
However, we found that in practice, this optimization generally requires a larger $k$ value to achieve chemical accuracy then simultaneous optimization of all $k$ sets of amplitudes in $k$-UpCCGSD. Therefore, for the results presented below, we optimized all $k$ sets of amplitudes simultaneously.
}

In general, with UCC methods it is not clear whether one obtains global minima
of the energy for a given class of wavefunctions.
Efficiently obtaining a global minimum in a non-linear optimization problem is an
open problem in applied mathematics.\cite{Nocedal2006}
In order to approximate the true minimum, each gradient-based optimization was therefore carried out
between thirty and two hundred times (depending on the cost) starting from
randomly chosen initial points.

\insertnew{
We note that the BFGS optimization as we have performed it here on a classical computer is unsuitable
for use on a quantum device due to the stochastic error associated with the
measurement of observables in the VQE framework. Given this, it will be necessary to
find better ways to handle optimization for large scale VQE experiments.
}
\subsection{Applications to Chemical Systems}

We now describe application of the three UCC ans{\"a}tze UCCSD, UCCGSD, and $k$-UpCCGSD,to three molecular systems possessing different geometries, namely \ce{H4}, \ce{H2O}, and \ce{N2}.

\subsubsection{\ce{H4}(in D$_\text{4h}$ and D$_\text{2h}$ symmetry)}
\ce{H4} is an interesting model system for testing CC methods with singles and doubles. We study here the potential energy curve of \ce{H4} for deviations from the square geometry with fixed bond distance, $R_\text{H-H} = 1.23$ \AA. Then we vary $R$ in the following coordinate system (values are given in \AA),
\begin{align*}
\text{H1}&: (0,0,0)\\
\text{H2}&: (0,0,1.23)\\
\text{H3}&: (R,0,0)\\
\text{H4}&: (R,0,1.23).
\end{align*}
This particular geometry setup has been used by others in Refs.~\citenum{Small2012}, \citenum{Paldus1993,Mahapatra1999,Kowalski2000,Jankowski1980,Evangelista2006}. At $R=1.23$ \AA\: (the D$_\text{4h}$ geometry), we have two quasidegenerate RHF determinants, which poses a great challenge to single-reference CC methods with only singles and doubles. 

We assess the ground state UCC methods including those developed in this work and compare them against RCCSD and coupled-cluster valence bond with singles and doubles (CCVB-SD) within the minimal basis, STO-3G.\cite{Hehre1969,Collins1976} CCVB-SD corrects for ill-behaving quadruples in RCCSD and is able to break any number of bonds exactly within the valence active space. In this sense, it is one of the most powerful classical CC methods with singles and doubles within the valence active space. There are two solutions for RCCSD and CCVB-SD, each one being obtained with one of the two low-lying RHF determinants. The two RHF solutions cross at $R=1.23$ \AA. We present the results obtained with the lowest RHF reference 
for a given $R$.

\begin{figure}[h!]
\includegraphics[scale=0.5]{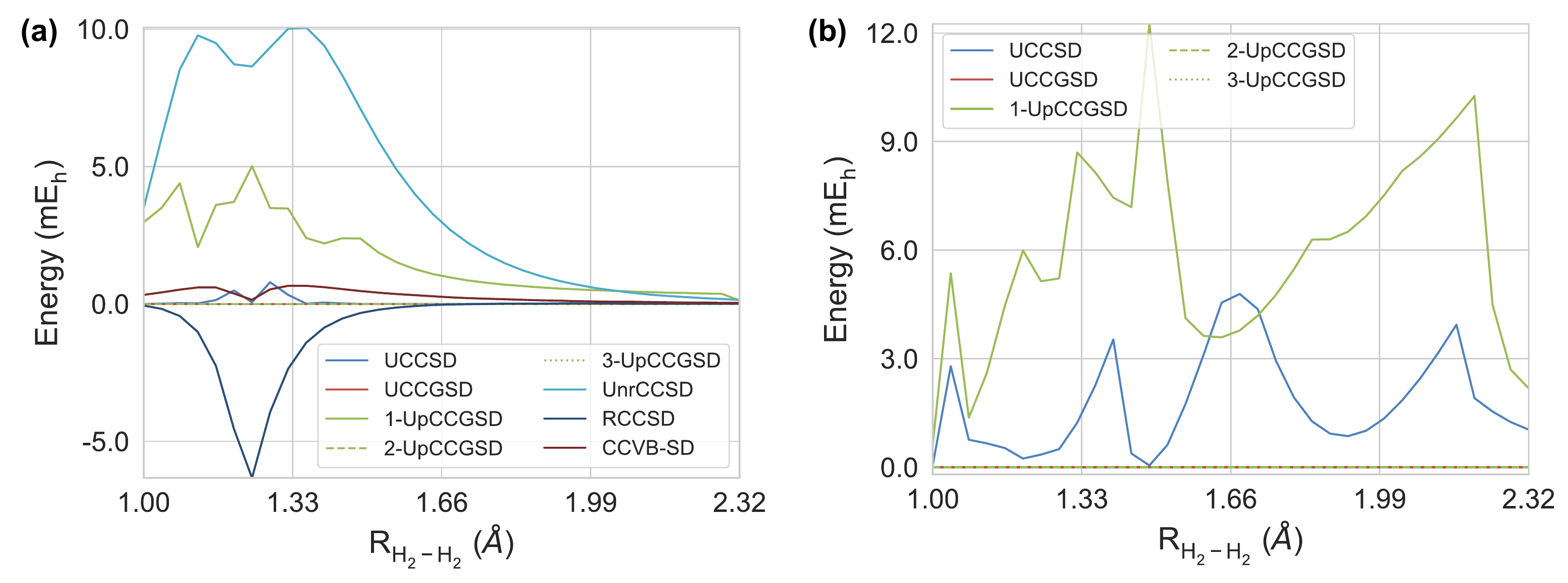}
\caption{\label{fig:h4sto3g}
The error in the absolute energy of the various CC methods examined in this work for (a) the ground state and (b) the first excited state of \ce{H4} as a function of the distance between two \ce{H2}'s. The basis set used here is STO-3G {(\(N=8\), \(\eta=4\))}. \revision{For both plots, UCCGSD, 2-UpCCGSD, and 3-UpCCGSD are overlapping near zero error in the absolute energy. }
}
\end{figure}

In Figure \ref{fig:h4sto3g} (a), we present the absolute energy error in ground
state of the aforementioned CC methods as a function of $R$. We first point out
that unrestricted CCSD (UnrCCSD) performs worst in an absolute sense among the methods examined here. This is because
the H-H distance in each \ce{H2} is stretched enough to get spin-contamination
on each \ce{H2}. This makes the entire potential energy curve of \ce{H4} heavily spin-contaminated
within the range of $R$ examined. RCCSD has clearly gone non-variational while
CCVB-SD remains above the exact ground state energy at all distances. Except
1-UpCCGSD and UCCSD, all the UCC variants are numerically exact. 1-UpCCGSD is
much worse than all the rest of UCC methods and adding one more product (i.e.
2-UpCCGSD) makes the energy numerically exact. 

Unlike full doubles CC
models, the energy of $k$-UpCCGSD is generally not invariant under unitary rotations among
orbitals. This is likely a primary cause of the multiple unphysical local minima observed for 1-UpCCGSD.
This problem can be ameliorated by increasing the value of $k$, as shown in Figure \ref{fig:h4sto3g} (a).
The difficulty of optimizing pair wavefunctions has been discussed in some earlier works.
Interested readers are referred to Ref.~\citenum{VanVoorhis2002}.

In Figure \ref{fig:h4sto3g} (b), the performance of UCC methods on the first
excited state of \ce{H4} was assessed within the OC-VQE framework. It is clear
that UCCSD and 1-UpCCGSD exhibit larger errors than those of the ground state.
This illustrates a potential drawback of OC-VQE in terms of accuracy when we do
not have a high quality ground state. However, with better ans{\"a}tze this drawback
can be made insignificant. The excited states from UCCGSD, 2-UpCCGSD, and
3-UpCCGSD are numerically exact, illustrating the power of these novel
wavefunction ans{\"a}tze which go beyond the capability of UCCSD \insertnew{while
also offering a lower asymptotic scaling}.

\begin{table}
\includegraphics[scale=0.50]{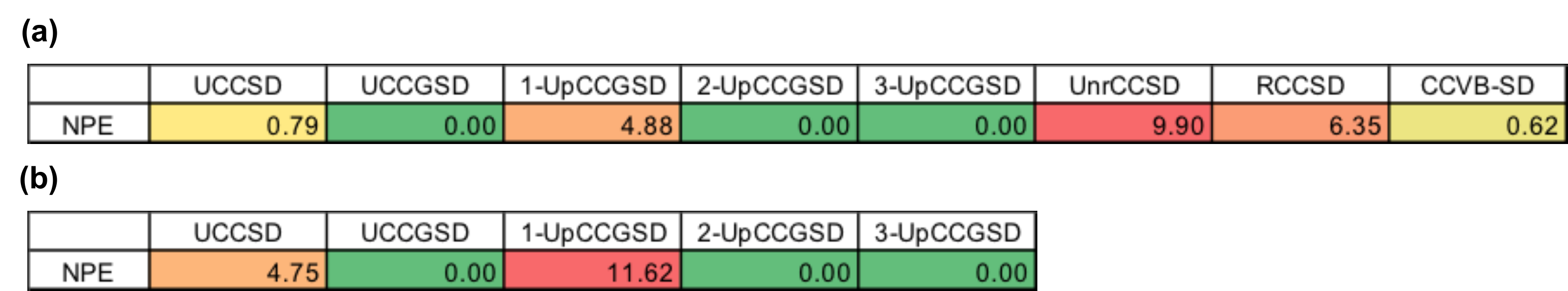}
\caption{\label{tab:h4sto3g}
The non-parallelity error (NPE) (m$E_h$) in (a) the ground state and (b) the first excited state of \ce{H4} within the STO-3G basis set {(\(N=8\), \(\eta=4\))}.
}
\end{table}

In Table \ref{tab:h4sto3g}, we present the non-parallelity error (NPE) in the
ground state and the first excited state for each CC method. 
NPE is defined as the difference between the maximum and minimum error and is a useful
measure of performance, since we are 
interested in relative energetics in most chemical applications.
In the ground state, UnrCCSD is the worst in terms of NPE. CCVB-SD is comparable to UCCSD and RCCSD and 1-UpCCGSD are comparable. UCCGSD, 2-UpCCGSD, and 3-UpCCGSD all have zero NPEs as they are numerically exact everywhere. In the case of the first excited state, UCCSD and 1-UpCCGSD performs worse than their ground state performance as observed before. All the other UCC methods are numerically exact.

We repeat the same calculations within the 6-31G basis. There are a total of 16 spin-orbitals in this case: in terms of resource on a quantum device this corresponds to the most expensive calculation reported in this work.
\joonho{}{This test is interesting because some dynamic correlation effects can be captured in 6-31G, in contrast to STO-3G, and these pose a greater challenge to pair CC methods.}
\begin{table}
\includegraphics[scale=0.41]{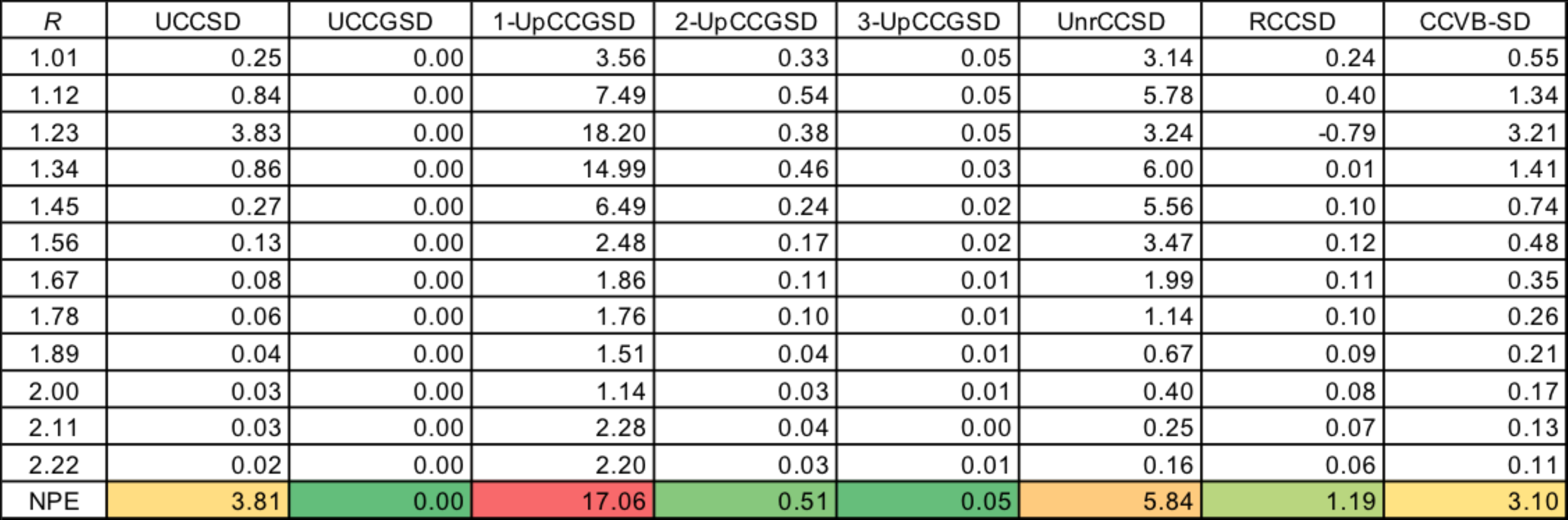}
\caption{\label{tab:h4631ga}
The error in absolute energy (m$E_h$) and non-parallelity error (NPE) (m$E_h$) in the ground state of \ce{H4} within the 6-31G basis {(\(N=16\), \(\eta=4\))} as a function of the distance ($R$) between two \ce{H2}'s (\AA).
}
\end{table}

In Table \ref{tab:h4631ga}, the error in the ground state is presented as a function of $R$. In terms of NPE, UCCGSD is again numerically exact and thus best. 2-UpCCGSD and 3-UpCCGSD are within 1 m$E_h$ of UCCGSD and exhibit larger errors than the corresponding results in the STO-3G basis.
RCCSD performs better with the 6-31G basis set and it is better than UCCSD. As it clearly becomes non-variational at $R = 1.23$ \AA, we suspect that this is a fortuitous outcome for RCCSD.
Moreover, UnrCCSD is the worst amongst the traditional CC methods considered in this work, which emphasizes the importance of spin-purity.

\begin{table}
\includegraphics[scale=0.41]{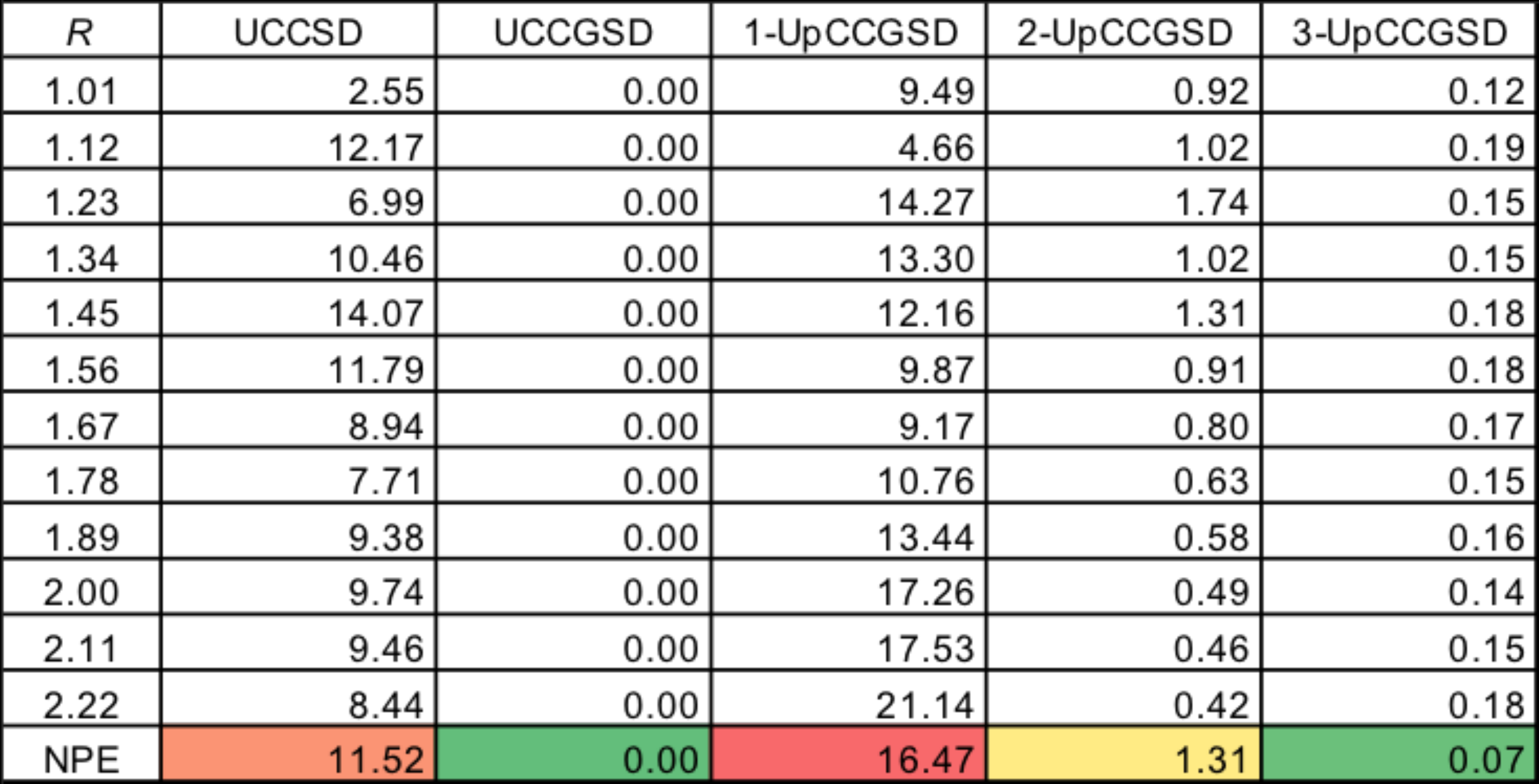}
\caption{\label{tab:h4631gb}
The error in absolute energy (m$E_h$) and non-parallelity error (NPE) (m$E_h$) in the first excited state of \ce{H4} within the 6-31G basis {(\(N=16\), \(\eta=4\))} as a function of the distance ($R$) between two \ce{H2}'s (\AA).
}
\end{table}

Lastly, we discuss the quality of the first excited state from UCC methods on
\ce{H4} within the 6-31G basis set \cite{Ditchfield1971} as presented in Figure
\ref{tab:h4631gb}. It is immediately obvious that the degraded ground state
performance of UCCSD is amplified in the excited state calculation and that
1-UpCCGSD continue to perform poorly. This is consistent with the STO-3G
results. However, it should be emphasized that UCCGSD is still numerically exact
and the 3-UpCCGSD error is still less than \(0.1\) m$E_h$. UCCSD's poor
performance strongly validates our development of better wavefunction
ans{\"a}tze beyond UCCSD, particularly for obtaining good excited states within the OC-VQE framework.

\subsubsection{Double Dissociation of \ce{H2O} (C$_\text{2v})$}
The double dissociation of \ce{H2O} is another classic test platform for various
wavefunction methods.\cite{Brown1984,Olsen1998,Li1998,Ma2006} As we stretch two
single bonds, we have total 4 electrons that are strongly entangled. The traditional RCCSD method can easily become non-variational, as will be demonstrated below. At a fixed angle $\theta_\text{HOH} = 104.5^\circ$ and within the C$_\text{2v}$ symmetry, we varied the bond distance between H and O and obtained potential energy curves for various CC methods within the STO-3G basis set.\cite{Hehre1969,Collins1976}

\begin{figure}[h!]
\includegraphics[scale=0.5]{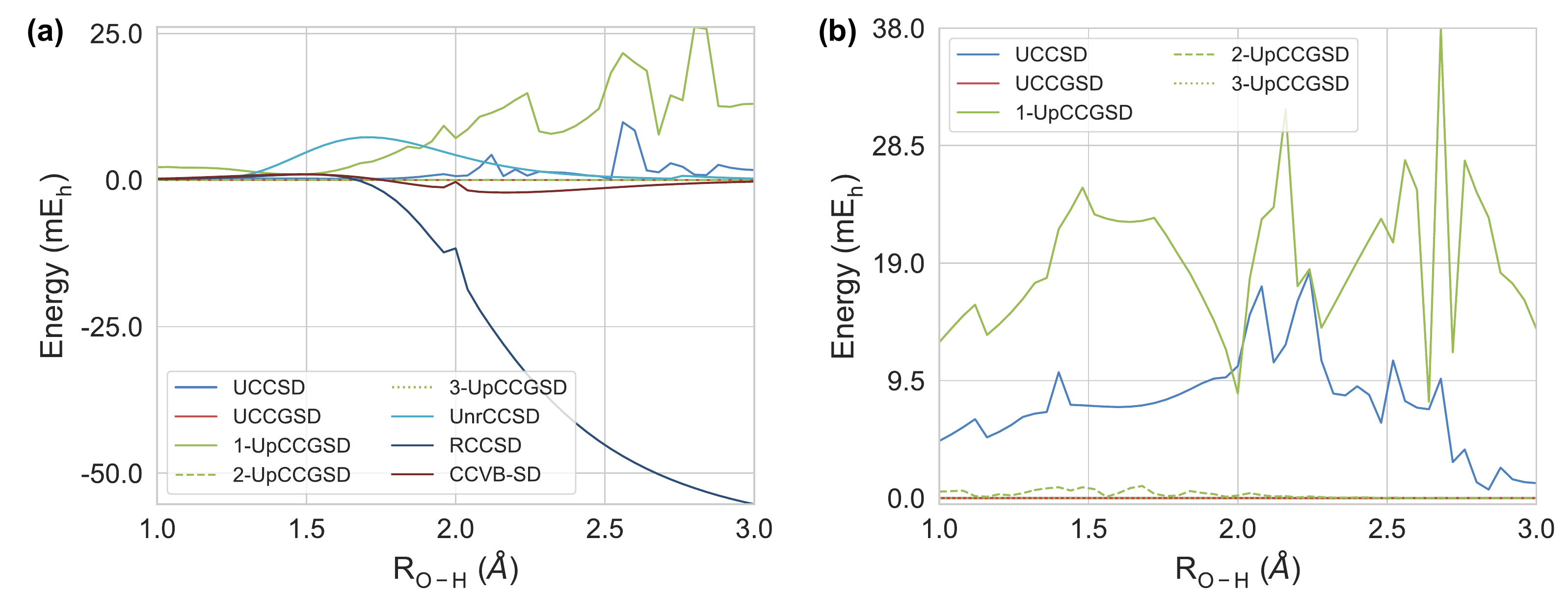}
\caption{\label{fig:h2o}
The error in the absolute energy of the various CC methods examined in this work for (a) the ground state and (b) the first excited state of \ce{H2O} as a function of the distance between \ce{O} and \ce{H}. The basis set used here is STO-3G {(\(N=12\), \(\eta=8\))}. \revision{For the ground state (a), UCCGSD, 2-UpCCGSD, and 3-UpCCGSD are overlapping near zero error in the absolute energy. For the excited state (b), UCCGSD and 3-UpCCGSD are overlapping near zero error in the absolute energy.}
}
\end{figure}

In Figure \ref{fig:h2o}, the error in the absolute energy of the ground state
and the first excited state of \ce{H2O} is presented as a function of the
\(R_\text{O-H}\) distance. In Figure \ref{fig:h2o} (a), RCCSD performs much worse than CCVB-SD and UnrCCSD especially after 1.75 \AA\: and exhibits a very significant non-variationality upon increasing the O-H distance.
\joonho{}{There is a kink between 2.02 \AA\: and 2.04 \AA\: in both RCCSD and CCVB-SD, that 
is due to a change in the character of the converged amplitudes. The RHF solutions for these CC calculations are delocalized and obey spatial symmetry. 
We also note that there is another spatially-symmetric RHF solution that is lower in energy than the orbitals we found. This solution starts to appear from 2.02 \AA\: and is more stable than the other for longer bond distances. This solution has orbitals either localized on O or two H's. This reference yields much higher CCVB-SD and RCCSD energies at 2.04 \AA. 
These two low-lying RHF solutions might cause multiple amplitudes solutions close in energy. We found that the largest $T_1$ amplitude of CCVB-SD is 0.28 at 2.02 \AA\: and 0.07 at 2.04 \AA.
This discontinuity does not appear with a larger basis set such as cc-pVDZ so it is likely an artifact of using a minimal basis.
With the delocalized RHF solution, CCVB-SD performs best among the classical CC methods examined here.}

UCCSD and 1-UpCCGSD perform much worse than the other UCC methods, as also
observed above in \ce{H4}. Other UCC methods are more or less numerically exact
on the scale of the plot. The performance of the first excited state as
presented in Figure \ref{fig:h2o} (b) is consistent with the ground state
performance. UCCGSD and 3-UpCCGSD are numerically exact and 2-UpCCGSD is within
1 m$E_h$ for all \(R_\text{O-H}\) values.
UCCSD and 1-UpCCGSD do not deliver reliable excited state energies.

\begin{table}
\includegraphics[scale=0.51]{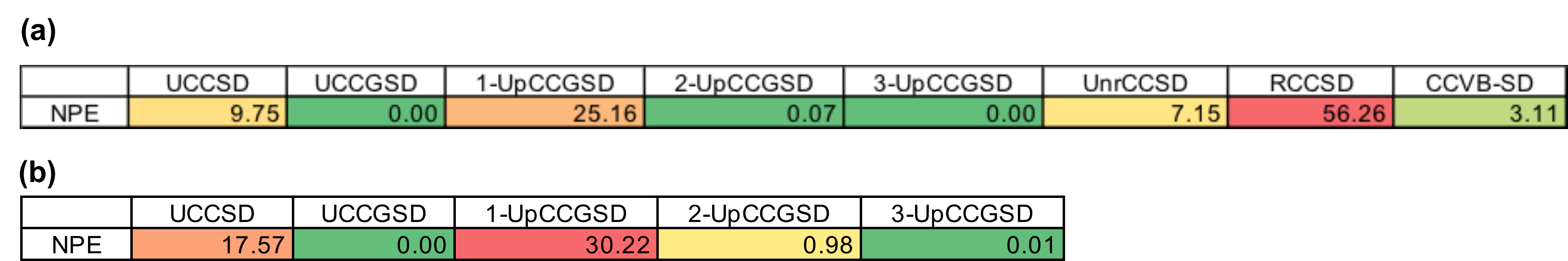}
\caption{\label{tab:h2o}
The non-parallelity error (NPE) (m$E_h$) in (a) the ground state and (b) the first excited state of \ce{H2O} within the STO-3G basis set {(\(N=12\), \(\eta=8\))}.
}
\end{table}

In Table \ref{tab:h2o}, we present the NPE of both the ground state in (a) and
the first excited state in (b) of \ce{H2O}. UCCGSD, 2-UpCCGSD, and 3-UpCCGSD all
yield reliable potential energy curves, while curves from the other methods are
not as reliable. It should be noted that UCCSD performs worse than the best
classical method considered here, UnrCCSD, but improved wavefunctions such as UCCGSD and 3-UpCCGSD are more or less exact for both states.

\subsubsection{Dissociation of \ce{N2}}
The dissociation of \ce{N2} is very challenging for CC methods with only singles and doubles.\cite{Siegbahn1983,Ma2006}
At a stretched geometry, there are a total of 6 electrons that are strongly
entangled.  RCCSD exhibits severe non-variationality and UnrCCSD has a
non-negligible non-parallelity error due to \joonho{}{poor performance in the intermediate bond length (spin-recoupling) regime.} To obtain a
qualitatively correct answer within the traditional CC framework with a RHF
reference, one would need RCCSD with the addition of triples, quadruples, pentuples and hextuples which contains far more excitations than RCCSD.
Alternatively, one could employ CCVB-SD as it is able to break \ce{N2} exactly within the STO-3G basis.\cite{Hehre1969,Collins1976}

\begin{table}
\includegraphics[scale=0.49]{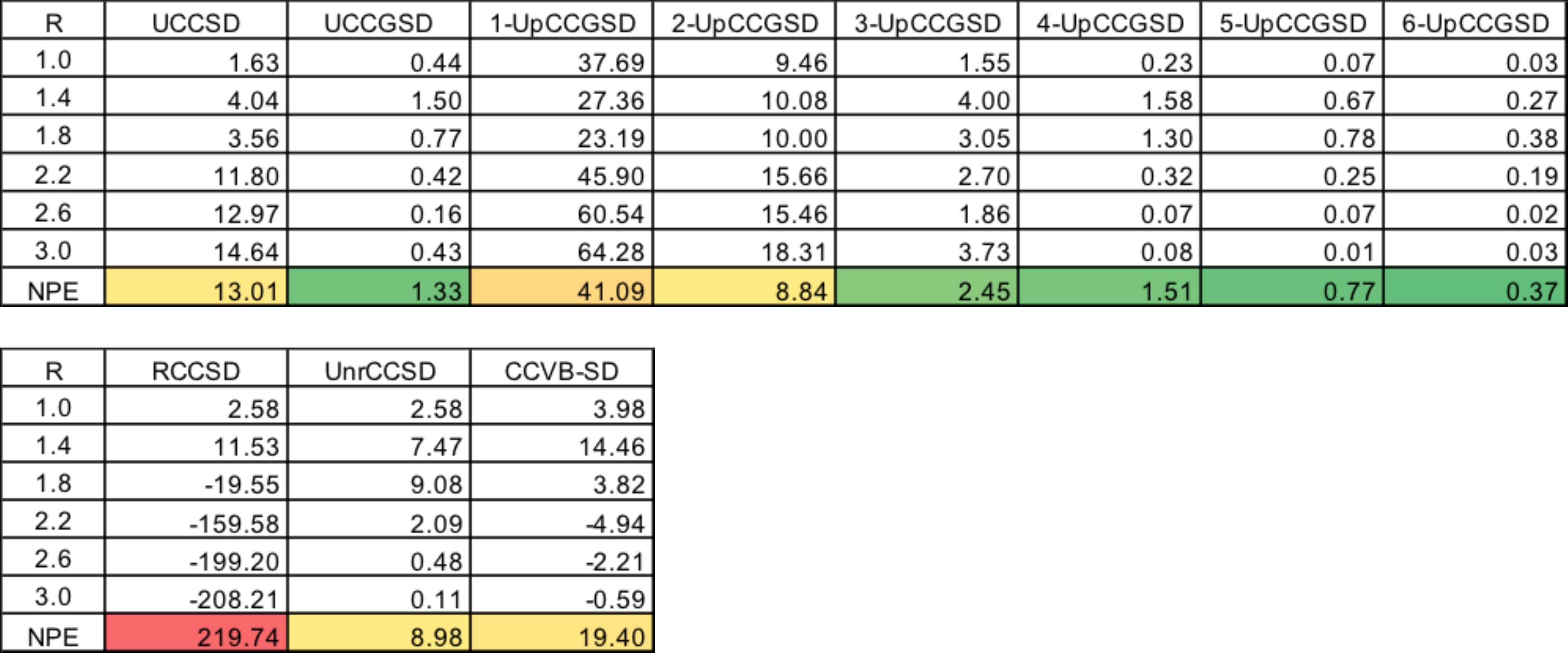}
\caption{\label{tab:n2a}
The error in absolute energy (m$E_h$) and non-parallelity error (NPE) (m$E_h$) in the ground state of \ce{N2} within the STO-3G basis {(\(N=16\), \(\eta=10\))} as a function of the distance ($R$) between two N's (\AA).
}
\end{table}

In Table \ref{tab:n2a}, we present the NPEs for ground state \ce{N2} for the various CC methods examined in this work.
In terms of the number of electrons that are strongly correlated, this system is the most challenging problem investigated in this work.
RCCSD is highly non-variational and not acceptably reliable for any distance considered
except for 1.0 \AA. CCVB-SD exhibits non-variationality but eventually dissociates properly.
However, in terms of NPE CCVB-SD is not reliable.
UnrCCSD has a NPE of 8.98 m$E_h$ due to poor performance at intermediate bond lengths.
UCC methods also struggle to properly dissociate. UCCSD is worse than UnrCCSD in terms of NPE. 
Furthermore, UCCGSD is now not numerically exact, with a NPE of 1.33
m$E_h$.
In order to achieve a NPE less than 1 m$E_h$, $k$ needs to be greater than 4. 
The fact that $k$-UpCCGSD is systemetically improvable and can achieve very
accurate results with a lower cost than UCCSD is very encouraging.

\begin{table}
\includegraphics[scale=0.51]{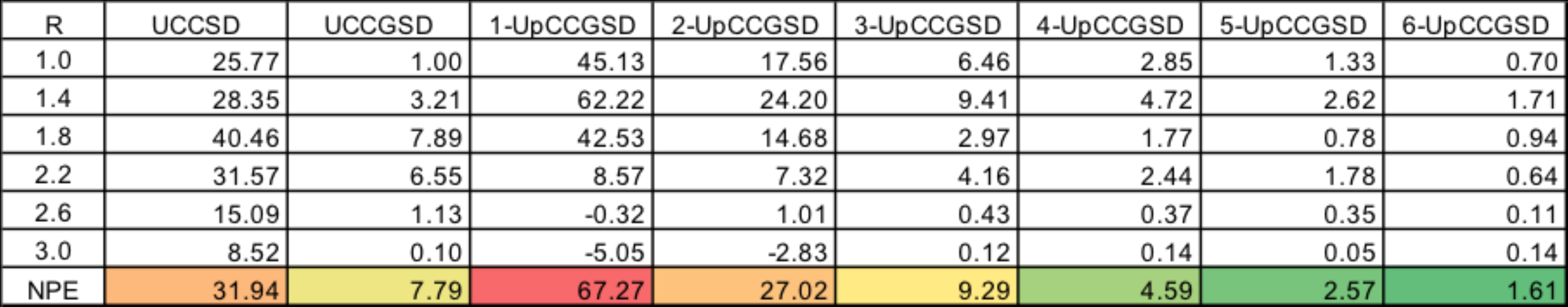}
\caption{\label{tab:n2b}
The error in absolute energy (m$E_h$) and non-parallelity error (NPE) (m$E_h$) in the first excited state of \ce{N2} within the STO-3G basis {(\(N=16\), \(\eta=10\))} as a function of the distance
($R$) between two N's (\AA).
}
\end{table}

Lastly, we discuss the performance of the UCC methods in the first excited state ${}^1\Pi_g$
which is presented in Table \ref{tab:n2b}.
Obtaining an accurate description for the first excited state of \ce{N2} within the OC-VQE framework is extremely challenging.
The best performing UCC method is 6-UpCCGSD with a NPE of 1.61 m$E_h$. UCCGSD
exhibits a NPE of 7.79 m$E_h$, which, while certainly better than that of UCCSD
(31.94 m$E_h$), is not close to the threshold for chemical accuracy.
These results highlight the challenge of constructing wavefunction ans{\"a}tze
capable of accurately representing the excited states of strongly correlated
systems.

\subsubsection{Discussion of Excited State Energies}
We analyze here the error of UCCGSD for the first excited state of \ce{N2} at 1.8 \AA\:, which is significant, at 7.89 m$E_h$.
For the purpose of demonstration, we ran another set of calculations with an exact orthogonality constraint constructed from the exact ground state. The results obtained with this exact constraint are presented in Table \ref{tab:exact}.
\begin{table}[]
\centering
\begin{tabular}{c|r|c}
\toprule
Determinants          & Error  	 &  Reference   \\ \hline
1                 & 10.23    &  Ground State RHF  \\
2                 & 3.18     &  Singly Excited Configuration ($\pi_x \rightarrow \pi_x^*$) \\
4                 & 0.45     &  Two Singly Excited Configurations ($\pi_x \rightarrow \pi_x^*$ and $\pi_y \rightarrow \pi_y^*$) \\ \bottomrule
\end{tabular}
\caption{The error in absolute energy (m\(E_h\)) for the first excited state of \ce{N_2} at 1.8
  \AA\: when using the exact ground state for the OC-VQE penalty term together
  with the UCCGSD ansatz and multiple reference states. 
  Here $\eta=10$ electrons in $N=8$ spin-orbitals.
  \label{tab:exact}
}
\end{table}

The ground state RHF determinant is likely to be a poor reference state for excited states. This is clearly demonstrated in Table \ref{tab:exact} with an error of 10.23 m$E_h$ in the case of the ground state RHF reference. The first excited state of \ce{N2} is a rather simple electronic state in the sense that it is mainly dominated by single excitations from the ground state wave function. At 1.8 \AA, these single excitations are mainly $\pi\rightarrow\pi^*$ and there are a total of two excitations like this along $x$ and $y$ cartesian components assuming that the molecular axis is the $z$-axis. Therefore, a more sensible starting point for OC-VQE would be to use these singly excited configurations. This leads to an error of 3.18 m$E_h$ with two determinants of the $\pi_x\rightarrow\pi_x^*$ type and to an error of 0.45 m$E_h$ with additional two determinants of the $\pi_y\rightarrow\pi_y^*$ type. A total of 4 determinants (or 2 spin-adapted singlet configurations) were enough to reach the chemical accuracy. In general, a much more sensible reference state for excited states like this can be cheaply obtained via regular linear response methods such as configuration interaction singles.\cite{Dreuw2005} Furthermore, the natural transition orbital basis \cite{Dreuw2005} can be used to generate a minimal multi-determinantal reference which will be usually of two determinants.

\subsubsection{Summary of Chemical Applications}
\begin{table}[h!]
\includegraphics[scale=0.6]{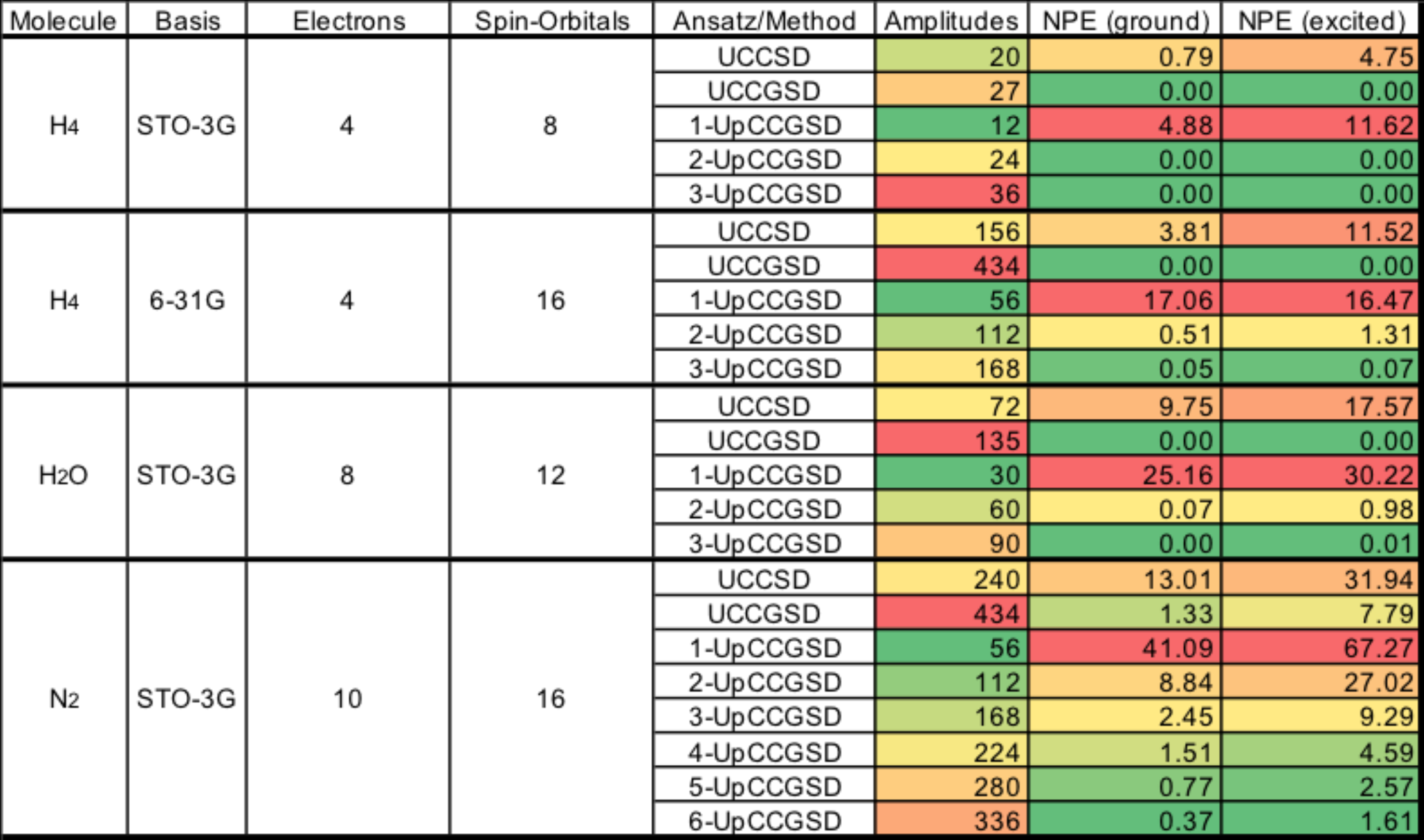}
\caption{\label{tab:final}
A summary of the results of this work: the number of amplitudes and the non-parallelity error (NPE) (m$E_h$) for each method applied to each molecule and basis. The excited NPEs are obtained with restricted Hartree-Fock references.
}
\end{table}
In Table \ref{tab:final}, we  present a summary of the results in this section. In particular, we focus on the tradeoff between the number of amplitudes and the accuracy (i.e. NPE). UCCSD does not perform very well given the number of amplitudes. $k$-UpCCGSD with a similar number of amplitudes always performs better than UCCSD which demonstrates the compactness of $k$-UpCCGSD. UCCGSD offers very accurate energies at the expense of \revision{requiring} a significant number of amplitudes. In all cases we considered it was possible to achieve chemical accuracy using $k$-UpCCGSD with less amplitudes than UCCGSD. We also note that excited states are in general more challenging than ground state calculations. \revision{Furthermore, there is no fortuitous error cancellation in excitation gaps in this approach. Therefore, it is important to obtain near-exact energies for both ground and excited states in order to achieve chemical accuracy for excitation gaps.} As noted above, using multi-determinantal reference wavefunction can improve the accuracy significantly. Considering the tradeoff between the cost and the accuracy, we recommend $k$-UpCCGSD for general applications. \joonho{}{However, it should be noted that for $k$-UpCCGSD to be effective, it is essential to choose $k$ large enough to obtain sub-chemical accuracy. Otherwise the lack of smoothness associated with this novel ansatz will inhibit application goals such as exploring potential energy surfaces.}

\section{Summary and Outlook}

In this work, we have presented a new unitary coupled cluster ansatz suitable for preparation, manipulation, and measurement of quantum states describing molecular electronic states, $k$-UpCCGSD, and compared its performance to that of both a generalized UCC ansatz UCCGSD, and the conventional UCCSD.  A resource analysis of implementation of these new
wavefunctions on a quantum device showed that $k$-UpCCGSD offers the best asymptotic scaling with respect to both circuit depth and amplitude count. Specifically, the circuit depth for $k$-UpCCGSD scales as $\mathcal O({kN})$ while that for UCCGSD scales as $\mathcal O({N^3})$ and that for UCCSD with $\mathcal O({(N-\eta)^2\eta})$. 

We performed classical benchmark calculations with these ans{\"a}tze for the ground state and first excited state of three molecules with very different symmetries, \ce{H4} (STO-3G, 6-31G), \ce{H2O} (STO-3G), and \ce{N2} (STO-3G), to analyze the relative accuracy obtainable from these ans{\"a}tze. Comparison was also with results from conventional coupled cluster wavefunctions where relevant.
The benchmarking calculations show that the new ansatz of unitary pair coupled-cluster with generalized singles and doubles ($k$-UpCCGSD) offers a favorable
tradeoff between accuracy and time complexity. 

We also made excited state calculations, using a variant of the recently proposed orthogonally
constrained variational quantum eigensolver (OC-VQE) framework \cite{higgott2018variational}. Our implementation of this takes advantage of the close
relation of this approach to some excited state methods in quantum chemistry.\cite{higgott2018variational, Cullen2011, Evangelista2013}
OC-VQE works as a variational algorithm where there a constraint in imposed on the
energy minimization in order to ensure the orthogonality of an excited state to a ground state wavefunction that has been previously obtained from a ground state VQE hybrid quantum-classical calculation. This approach requires only a
modest increase in resources to implement on a quantum device compared to the resources required for ground
state VQE, and is furthermore capable of targeting states outside of a small linear response subspace defined from the VQE ground state. 

Assessing the classically computed potential energy curves of these three molecules, we found that the
error associated with excited states obtained by the OC-VQE approach in conjunction with the standard UCCSD reference, is considerably larger than the error of the ground state calculation. The excited states of UCC singles and doubles 
are never of high quality, except for simple two-electron systems where UCCSD is exact.\cite{higgott2018variational}
We found that energies of both ground and excited states can be greatly improved by employing 
either UCCGSD, i.e., UCC with generalized singles and doubles, or the $k$-fold products of $k$-UpCCGSD. Furthermore, we demonstrate that the quality of excited state calculations in the OC-VQE framework can be dramatically improved by choosing a chemically motivated reference wavefunction.

UCCGSD was found to be numerically exact for \ce{H4} (STO-3G, 6-31G) and \ce{H2O} (STO-3G) for both ground and excited states. 
However, its non-parallelity error (NPE) is 1.33 m$E_h$ for the ground state of \ce{N2} and 7.79 m$E_h$ for the first excited state of \ce{N2}.
$k$-UpCCGSD was found to be numerically exact for a large enough $k$, where the required value of $k$ increases with the difficulty of the problem. It would be interesting to study the required value of $k$ for fixed accuracy on a broader class of problems in the future.

In summary, this work demonstrates the advantages of wavefunction ans{\"a}tze
that go beyond UCCSD and indicates the desirability of further refinement of such ans{\"a}tze to forms that are accurate for both ground and excited states. The performance of $k$-UpCCGSD is particularly encouraging, showing a tradeoff between accuracy and resource cost that allows chemical accuracy to be achieved with resources scaling only linearly in the number of spin-orbitals.
Our analysis of excited states indicates that these pose significant challenges and there is a need for focus on these.  In particular, 
we anticipate that further development of novel algorithms not within the variational framework may be necessary to obtain high quality excited
state energetics, particularly when working with an approximate ground state.

Finally we note that the wavefunctions we have investigated in this work can
be fruitfully combined with existing classical approximations to UCC based on the
truncation of the Baker-Campbell-Hausforff expansion of \(\ev{e^{T^\dagger
    -T}He^{T-T^\dagger}}{\phi_0}\). \cite{Watts1989,Bartlett1989,Kutzelnigg1991} This would allow for the efficient
initialization of the cluster amplitudes, making it possible to further optimize
them using the VQE hybrid approach to quantum computation, and also avoiding the difficulties posed by a random initialization.\cite{mcclean2018barren} 
{In future work, it would be interesting to further explore the balance between the cost and accuracy of unitary coupled cluster ans{\"a}tze obtained here by building on chemically motivated approximations. Two especially promising directions that we believe could yield a further reduction of the number of amplitudes and the gate depth required for a fixed accuracy, are i) the adaption the recently proposed full coupled-cluster reduction \cite{Xu2018} method for use on a quantum computer, and ii) the elimination of singles amplitudes through the use of approximate Br{\"u}ckner orbitals \cite{Dykstra1977,Handy1989,Krylov1998,Lochan2007,Leeoomp2} obtained by classical pre-processing.}
Ultimately, the resulting wavefunctions could
themselves serve as inputs to a fully quantum computation of more accurate
ground and excited state energies, e.g., with the quantum phase estimation algorithm, or to a quantum simulation of quantum dynamics.

\section{Acknowledgement}
The work of W. Huggins and K. B. Whaley
was supported by the U.S. Department of Energy, Office of Science, Office of Advanced Scientific Computing
Research, Quantum Algorithm Teams Program.
J. Lee and M. Head-Gordon acknowledge support from the Director, Office of Science, Office of Basic Energy Sciences, of the U.S. Department of Energy under Contract No. DE-AC02-05CH11231.
We thank Dr. Dave Small for stimulating discussions on the performance of CCVB-SD on the \ce{H2O} dissociation benchmark.

\section{Table of Contents Graphic}
\begin{figure}[h!]
\includegraphics[scale=0.35]{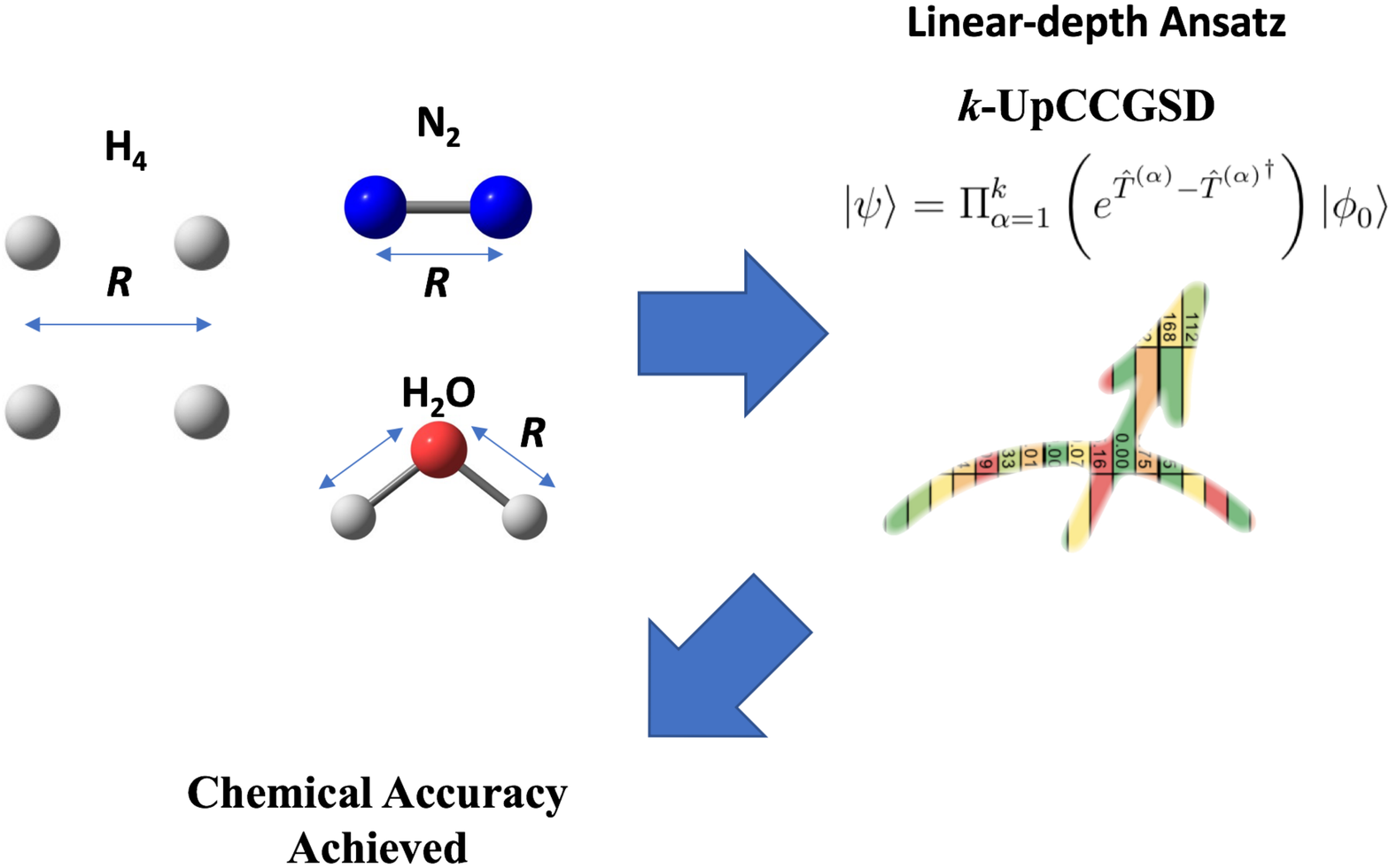}
\end{figure}

\bibliography{refs}{}
\end{document}